\documentstyle[12pt]{article}

\begin{document}

\def\eq{\begin{equation}}
\def\en{\end{equation}}

\begin{titlepage}

\vskip .27in
\begin{center}
{\large \bf {Real-world options: smile and residual risk}}

\vskip .27in
Jean-Philippe Bouchaud$^{1}$, Gulia Iori$^{2}$ and Didier Sornette$^{3}$

{\it 1) Service de Physique de l'Etat Condens{\'e}, CEA-Saclay, 91191 Gif
sur Yvette
CEDEX, France}
\vskip .2in
{\it 2) Service de Physique Th\'eorique, CEA-Saclay, 91191 Gif
sur Yvette
CEDEX, France}
\vskip .2in

{\it 3) Laboratoire de Physique de la mati\`ere Condens\'ee, Universit\'e
de Nice-Sophia
Antipolis, B.P. 70, Parc Valrose, 06108 Nice CEDEX 2, France. }

\vskip .2in

\end{center}
\vskip 2cm
\begin{center} { \bf Abstract } \end{center}
\vskip 1cm

We present a theory of option pricing and hedging, designed to address
non-perfect arbitrage, market friction and the presence of `fat' tails. An
implied volatility `smile' is predicted. We give precise estimates of the
residual risk
associated with optimal (but imperfect) hedging.

\vskip 1cm
Accepted for publication in Risk Magazine (December 1995).
\end{titlepage}

\section{Introduction}

In 1973, Black and Scholes \cite{BS} developed the mathematical theory
of perfectly {\it hedged} options in an {\it arbitrage-free}
lognormal
random walk asset price model. Fundamental for the mathematical formulation
and deeply
rooted in the mind of both theoreticians and practitioners, evolved the
notion that a
general strategy of pricing and coverage of derivatives
involves the {\it hedging away of all risks}. However, we do not live in a
Black-Scholes world
: \begin{itemize}

\item asset prices do not follow a continuous time lognormal processes, but may
exhibit large
jumps and be distributed with fat tails, in addition to present subtle
correlations
between increments at different times (the markets are `incomplete'),

\item the replicating portfolio cannot be implemented exactly, since it
involves
continuous rebalancing: several `imperfections' such as transaction costs,
delays and lack of liquidity necessarily introduce a finite time scale for
transactions.

\end{itemize}

Confronted with these practical problems, many authors have attempted to
generalize the Black and Scholes strategy. Here, we present an intuitively
appealing
formalism which is flexible enough to address these problems efficiently
and show how
it can be implemented practically. Our method is based on a global
(integral) representation of wealth
balance, in contrast to Black-Scholes local (differential) approach. The
mathematical tool-box
is that of functional integration and derivation, in contrast to the
standard It\^o calculus and partial
differential equation formalism.
We propose two main new concepts :
\begin{itemize}
\item  The
minimization of the {\it non zero residual risk} as a criterion to fix an
optimal
strategy \cite{BS94}. One should note in this respect that somewhat related
ideas have previously been proposed in the mathematical literature in a rather
formal way \cite{Others}. Furthermore, our hypothesis are more general and the
resulting formulae are easily implemented numerically.
\item For
strongly fluctuating assets, option pricing and its hedging is obtained by
a `tail chiseling' of
wealth distributions as a treatment of large risks \cite{BS95}.
\end{itemize}

One of our
main thrust is to
demonstrate how these ideas can be put in practical terms, using relatively
simple mathematics.

Let us discuss it for the basic European call option pricing problem  --
generalisation to more complicated products does not
involve any conceptual difficulties.  Suppose that an operator wants to buy
a given
share, a certain time $t=T$ from now ($t=0$ at which the current value is
$x(t=0) \equiv x_0$),
at a fixed
`striking' price $x_c$. If the share value at $t=T$, $x(T)$, exceeds $x_c$,
the operator `exercises' his option, with an immediate profit
difference $x(T)-x_c$. On the contrary, if $x(T) <  x_c$ the operator may
not buy the
share. What is the price  ${\cal C}(x_0,x_c,T)$ of this possibility given to
the operator by -say- the ``bank'', and what trading strategy should be
followed by the
bank between now and $T$, depending on what the share value $x(t)$ actually
does
between $t=0$ and $t=T$? These are the two questions answered by Black and
Scholes in the context of an ideal complete market,
and which we wish to consider from a somewhat different point of view adapted
to real-world markets.

In common with Black and Scholes, the starting point is to write a wealth
balance for the bank. However, instead of computing the {\it instantaneous}
variation of
the value of its portfolio (which assumes continuous time), we write a global
balance at time $T$, and assume that time is discrete, with a certain
`microscopic' time scale $\tau$ below which trading is impossible (or, in
the presence of transaction costs, very unfavorable -- see below).

We shall assume for simplicity that the risk-free interest rate $r$ is
constant; more involved developments (including a random interest rate) are
reported elsewhere. The {\it global}
balance $\Delta W$ for the bank between $t=0$ and $t=T$ is thus given by:
 $$
\Delta W =  {\cal C}(x_0,x_c,T)\exp(rT)  -
\max(x(T)-x_c,0) +$$
$$+ \sum_{t=0}^T \ \phi(x,t)\exp(r(T-t)) [\Delta_\tau x -rx(t)]  \eqno (1) ,$$
with $x_c$ the striking price, $x_0$ the initial ($t=0$) share  price, and
$\phi(x,t)$
is the number of shares owned by the bank at time $t$, knowing that the
share price is
$x$. $\Delta_\tau x$ is the difference of share values between times
$t+\tau$ and $t$.
${\cal C}(x_0,x_c,T)$ is the looked for option price. Expression (1), which
in the continuous
case is in fact the time-integrated version of Black and Scholes's
differential equation,
has a very
intuitive meaning :
\begin{itemize}
\item The first term is the gain from
pocketing from the buyer the option price at $t=0$, discounted at time $T$.
\item The second term gives the potential loss equal to $-(x(T)-x_c)$
if $x(T) > x_c$ (i.e. if the option is exercized) and zero otherwise.
\item The third term
quantifies the effect of the trading between $t=0$ and $t=T$: the true
variation of
wealth $W$ between $t$ and $t+\tau$ is only due to the
fluctuations of the share price, i.e. $\phi (x,t) \Delta_\tau x $, corrected by
the fact that $x \phi(x,t)$ has not benefited from the risk-free interest rate.
(Note that the term $x {\Delta_\tau \phi } $ describes conversion of shares
into other
assets or the reverse, but not a real change of wealth).
\end{itemize}

The bank wealth variation $\Delta W$ depends a priori on the specific
realization
$\{x(t)\}|_{t=0-T}$ of the asset price. For a continuous log-normal
process, the result of Black and Scholes is that $\Delta W$ is strictly
vanishing for $\tau \rightarrow 0$ for a
particular choice of $\phi(x,t)$ and ${\cal C}(x_0,x_c,T)$.
Equivalently, both equalities $\langle \Delta W \rangle =0$ and $\langle
(\Delta
W)^2\rangle =0$ hold, where
$\langle ...\rangle $ denotes the average over all possible different
realisations of the history
$\{x(t)\}$. In other words, the averages $\langle ... \rangle$ are taken
over the initial
historical measure in which the drift is not in general equal to the
risk-free interest rate.
The first condition $\langle \Delta W \rangle =0$ is the `no free lunch'
condition. The second
expression  simply writes that the square of the bank wealth volatility
$\langle \Delta W^2 \rangle $ is
zero which ensures the hedging away of all risk.
This is just another view point to retrieve Black and Scholes results,
holding true for any quasi-Gaussian, time-continuous processes \cite{BS94}.

\vskip 1cm
\section{Risk-corrected option prices}

In the real world where $\Delta W$ cannot be made to vanish exactly, we
propose to apply the
spirit of Markowitz portfolio approach to the total bank wealth balance,
viewed as an
effective portfolio estimated at the time $T$ of the death of the option.
In this goal, one
has to calculate the average return $\langle \Delta W \rangle $ and the
second moment
(related to the square of the bank
wealth volatility) $\langle \Delta W^2 \rangle $ when they exist (the case
of very strongly fluctuating `L\'evy
processes'
with ill-defined theoretical volatility will be addressed below).
For the time
being, we assume that both  $\langle \Delta W \rangle $ and $\langle \Delta
W^2 \rangle $ exist. The
optimal
strategy $\phi^*(x,t)$ that the bank must follow should certainly be such
that the uncertainty on the outcome is minimum. The reason
for this
is that the bank will adjust the option price according to its risk
aversion, for
instance with the criterion $$\langle \Delta W \rangle |_{\phi=\phi^*} =
\lambda \sqrt{\cal
R^*},
\eqno(2)$$
where ${\cal R}^* \equiv \langle \Delta W^2 \rangle -\langle \Delta W
\rangle ^2 |_{\phi=\phi^*}$
is the square of the volatility of the bank wealth. Eq. (2) simply means that
the bank wishes to keep the probability of global loss incurred by
delivering the
option at a certain level -- say $10 \%$. The minimisation of
${\cal R}^*$ is thus necessary to keep the option price itself as low
as possible. Alternatively, the knowledge of $\langle \Delta W \rangle
|_{\phi=\phi^*}$
and $\sqrt{\cal R^*}$ allows one to extract from the market price of a
given option the value of its
`$\lambda$' (a yet unused Greek letter).
$\lambda$ can be thought of as an objective (dimensionless) measure of the
real option price, in units of the real risk associated to the underlying asset
-- high $\lambda$'s corresponding to
expensive options. Note that the term in the right hand side of eq.(2)
allows the bank to introduce
a bid-ask offer range around a central `fair' price which itself is
independent of $\lambda$. This bid-ask offer range is
dependent upon the bank portfolio and its risk-aversion.

\subsection{Independent increments of the share value}

Let us suppose that the local slopes $\Delta_\tau x$ are
statistically independent (but not necessarily Gaussian) for different times.
In this case, both $\langle \Delta W \rangle$ and the `risk' measured as ${\cal
R}[\phi(x,t)]=\langle (\Delta
W)^2\rangle -\langle \Delta W \rangle ^2$ can be explicitely calculated. We can
then  determine the optimal strategy $\phi^*(x,t)$ through a
`functional minimisation':

$${\partial {\cal
R}[\phi(x,t)] \over \partial \phi(x,t)} |_{\phi=\phi^*}= 0.  \eqno(3)$$

from which a general expression of both $\phi^*(x,t)$ and of the residual
risk ${\cal R}^* = {\cal
R}[\phi^*(x,t)]$ can be derived \cite{BS94} -- see the `Technical sheet'
below).
 These results are valid for an
{\it arbitrary}
stochastic process with uncorrelated increments, including `jump', or
discrete-time, processes. The
process can furthermore be explicitely time-dependent, with a variance
which is a function of time, as
for the much studied `ARCH' processes,
which are Gaussian processes with a t-dependent variance ${\cal D}(t)$.

The obtained formulae can be simplified in several cases \cite{BS94} and allows
 us to retrieve Black and Scholes' result when $\Delta_\tau x$ are Gaussian
variables and $\tau$ tends to zero (continuous process). In particular, one
finds indeed  that ${\cal R}^*$ vanishes exactly in the continuous time
limit. One may also check that
in that case, the results on the option price and the optimal strategy are
indeed {\it
independent} of the average return of the share.

The condition ${\cal R^*}=0$ is very specific to the Gaussian can and does {\it
not hold} for a
more general stochastic process. This is the main difference between the
 present approach and
that of Black and Scholes and subsequent workers (see however \cite{Others}):

1) we
find that a vanishing
residual risk cannot be achieved in the general case and are able to
quantify it precisely (see Fig 2 below);

2) however, this does not imply that an optimal strategy does not exist. We
have indeed found an optimal $\phi^*(x,t)$ which minimize the risk
and which is a natural generalization of Black and Scholes result.

Note finally that other definitions of the risk are possible, for example
through higher moments of the distribution of $\Delta W$. The functional
minimisation technique presented here can easily be adapted
to these cases
-- although the
calculations are more cumbersome.

\subsection{Illustration: options on the MATIF}

These findings are
relevant to various concrete situations. In particular, strong deviations
from a Gaussian
behaviour (``leptokurtosis'') are often observed in many situations.
In this case, the above method allows one to estimate quantitatively
the residual risk
and correct the
trading strategy and the option price accordingly. We present an
illustration by comparing
the option price obtained from the proposed functional quadratic minimization
procedure and from Black and Scholes formula applied on MATIF options, with
a 30 days
maturity ($T=\frac{30}{365}$ days). In this goal, we have determined the
historical conditional distribution
$P(x',t'|x,t)$ from the MATIF daily quotation variations in the period
1990-1992. We have tested a certain degree of stationarity of the data by
constructing the distribution of price increments over various time
intervals. We then determined
$P(x',t'|x,t)$ by scanning the data set and using the assumption of
stationarity implying that $P(x',t'|x,t) = P(x'-x,t'-t)$.
In order to apply the Black and Scholes results,
we have fitted  $P(x'-x,t'-t=1)$ by a log-normal distribution, finding a
volatility
$\sigma^2 \simeq 10^{-5}$ per day.  $P(x'-x,t'-t=1)$ and its log-normal fit is
shown in Fig.1.
We can also apply
 directly our formulae using the empirically determined distribution
and obtain the option price ${\cal
C}(x_0,x_c,T)$ as a
function of the striking price $x_c$ (Fig 2) in a risk insensitive world
($\lambda=0$), and
compare it to the Black-Scholes price with the historical volatility
(dotted curve).
As is well known, this procedure underestimates the `true' price: the
presence of
`tails' in the distribution induces a larger effective volatility. One can
in fact
invert the Black-Scholes formula and determine an implied volatility from our
determination of the price. The result is given in the inset of Fig 2 and
reproduces the well known `smile': the effective volatility is stronger for
`out of the money' options. More importantly, the residual risk  ${\cal R}^*$
is
{\it non-zero}, it furthermore depends on the trading `frequency', i.e. the
number of
`inactive' days $\tau$ without rehedging. As expected, the larger this
number of days,
the  larger the risk : we have plotted in Fig 3 the quantity $\sqrt{{\cal
R}^*} $ as a
function of $\tau$, together with the total transaction costs $\cal K$
associated with
these rehedging, assuming that each trading has a cost equal to $0.05 \%$ of
the
share's value.  The full curve represents the sum of these two sources of
extra-cost: $\overline {\cal K}= \sqrt{{\cal R}^*}+{\cal K}$ is the
risk-aversion cost
($\lambda=1$) plus the transaction costs. Interestingly, $\overline {\cal
K}$ has a
{\it minimum}, which is the optimal trading rate ($\tau \simeq 10$ days for
this particular choice of parameters). Due to the impact of transaction costs,
it appears that a daily rehedging strategy is not reasonable -- unless
$\lambda$ is very
large (strong risk aversion). Note that $\overline {\cal K}$ represents an
appreciable fraction of the `fair' price.

Finally, we have compared in Fig 3 (Inset) the residual risk for our
optimal strategy with the residual risk obtained following the Black-Scholes
strategy. As expected, the latter is larger: the risk can be {\it substantially
reduced}
using our `optimal' strategy.

\subsection{Rare events and fat tails}

Our method can be adapted to many situations: more exotic options, or more
complicated stochastic processes, such as correlated Brownian motions
\cite{BS94} or
L\'evy processes, which have been
argued by many authors \cite{Levy} to be adequate models for short
enough time lags, when the kurtosis is large. The latter case is
interesting since
the notion of variance becomes ill-defined due to the presence of extremely
large fluctuation
(crashes). The above criterion fixing the optimal strategy $\phi(x,t)$,
based on a minimization of the
variance is thus meaningless. Intuitively, this comes from the
fact that the
variance is dramatically sensitive to
large price
variations. This indicates that moments are not sufficient anymore to capture
the informations
contained in the price and wealth distribution : one must study directly
the distributions
themselves, and more precisely the tails of the distribution $P(\Delta W)$
of the wealth variation $\Delta W$. By the laws of composition of L\'evy
laws, one can show that $P(\Delta W)$ decays as
 $W_0^\mu[\phi(x,t)] \over |\Delta W|^{1+\mu}$ for large losses ($\Delta W
\longrightarrow -\infty$),
 with $W_0$ depending
on $\phi(x,t)$.  We then propose to determine the optimal strategy
$\phi^*(x,t)$ by the
condition ${\delta W_0 \over \delta \phi(x,t)}=0$ -
corresponding to a minimization of the `catastrophic' risks, since
$W_0$ controls the {\it scale} of the distribution of losses, i.e. their order
of magnitude.

The full derivation of the solution of this minimisation problem will be
presented elsewhere; here we restrict to the simple `variational' ansatz where
$\phi(x,t)$ is taken to be a constant $\phi$, independent of $x$ and $t$, to be
optimized. Using Eq.(1), we need to calculate the distribution $P(\Delta
W)$ of large
losses. $\Delta W$ can
take large negative values when
\begin{itemize}
\item $x(t)$ drops dramatically : the option is not realized
($\max(x(T)-x_c,0)=0$) but the
bank looses $(x(T) - x_0) \phi$ due to its hold position. Taken alone, this
favors
$\phi\rightarrow 0$.
\item $x(t)$ increases much above $x_c$ : the bank has to produce the
share and thus loses $-(x(T)-x_c)$ but partially compensates this loss by
its holding of
$(x(T) - x_0) \phi$, thus resulting in a net loss of $-(1-\phi) (x(T) -
x_0)$. Taken alone,
this situation favors $\phi\rightarrow 1$.
\end{itemize}
There is thus a trade-off between the two possibilities, leading to a
non-trivial optimal
$\phi$.

In order to obtain it, we write the distribution $P(\Delta W)$ of large losses
as the
sum of these two
independent processes and get
$$W_0^\mu =(1 - {\cal P}) C_- \phi^{\mu} + {\cal P} C_+
(1-\phi)^{\mu} .\eqno(4)$$
where $C_{\pm}$ describe the `tails' of the historical distribution:
$$ P(x,T|x_0,0) \simeq_{x-x_0 \rightarrow \pm\infty}
\frac{C_{\pm}}{|x-x_0|^{1+\mu}}.$$ $\cal P$ is the probability at $t=0$
that the option will be
exercised. Minimisation of $W_0$ with respect to $\phi$ yields
$$
\phi^* = { [C_+ {\cal P}]^\zeta \over [C_+ {\cal P}]^\zeta+ [C_- (1-
{\cal P})]^\zeta } \qquad \zeta \equiv {1\over \mu-1}
\eqno(5)$$
A more complete treatment with an arbitrary $\phi(x,t)$,
in fact leads to
the same result given by eq.(5), but with a {\it time dependent} ${\cal
P}=\int_{x_c}^{\infty} dy'
P_0(y',T|y,t)$, which is the probability at time
$t$ at which the stock price is $y$ that the option will be exercized (at $T$).

If $(1 - {\cal P})C_- \ll  {\cal P}C_+$ (resp. $(1 - {\cal P})C_-\gg  {\cal
P}C_+$),
$\phi \rightarrow 1$ (resp. $0$), as is natural since
then one of the two histories dominate. If both have the same weight, the
intuitive result $\phi^* = {1 \over 2}$ is recovered. The resulting optimal
`scale'
$W_0(\phi^*)$ is now the correct measure of the risk which must be added to
the risk
neutral price with a coefficient depending on risk aversion.

The case where $\mu <  1$, for which the mean becomes itself ill-defined,
has been discussed in \cite{BS94}. Let us simply mention that the optimal
strategy in that case is either $\phi^*= 0$  or $1$. Finally,
 in the border case $\mu=2$ (asymptotically attracted to the `Gaussian' stable
law),
with symmetric tails $C_+=C_-$, we recover $\phi^*(x,t) = {\cal P}(x,t)$, which
is precisely Black and Scholes's result.

\section{Conclusion}

We have thus proposed a pragmatic and flexible theory of option pricing,
which naturally generalizes the Black-Scholes results to intrinsically risky
markets. Our method allows one
to decompose in readable way the option price into an irreducible part
corresponding
to a `fair game' condition {\it plus an extra-cost} $\overline {\cal K}$ which
includes the
residual risk associated with trading and transaction costs. The risk part is
minimized in a variational way with respect to the strategy. A well defined
trading
frequency appears, as a trade-off between risk and transaction costs. Our
precedure
is furthermore relatively easy to implement numerically, and we hope that
it will
prove useful to professionals.

\vskip 0.5cm

We wish to thank J.P. Aguilar, L. Mikheev, J. Miller and Ch. Walter for
interesting
discussions.

\vskip 1cm
{\bf Figure Captions}.
\\
\\
Fig. 1. :
Historical distribution of the daily relative variation $\Delta x \over x$ for
the MATIF daily quotation variations in the period
1990-1992 and comparison with the best fit to a normal distribution (dotted
curve). As is by now familiar, the empirical distribution has clearly visible
`fat tails' and a sharper maximum. For example, the kurtosis of the
distribution is equal to $28.1$ instead of $3$ for a normal distribution.

Fig.2 :
a) Comparison between the option price obtained from Eq.(8) with
$\lambda=0$ and assuming independent
increments (full line) and from the Black and Scholes formula with the
historical volatility (dotted
line), applied on MATIF options with 30 days maturity (with striking price
higher than the daily quotation
$x_0=100$) and put (with striking price lower then $x_0=100$) options. As
well known, Black and Scholes
formula underestimates option prices when using the historical volatility.
. The difference between the two curves
is  significantly larger than
the error bars from our numerical implementation of the analytical formula
based on historical data.

b) Inset: `Implied volatility' obtained by inverting the
Black-Scholes formula. The well known `volatility smile' is reproduced.
\\
\\
Fig 3. :
a) Risk aversion cost $\sqrt{{\cal R}^*} $ (dashed line) and
total transaction cost ${\cal K}$ (dotted line) associated with rehedging
as a function of
trading  time $ \tau $, for a 30 days option with $x_c=x_0$.
We assume that
trading costs $0.05 \%$ of the `share' value.
The sum of the two curves (full line) represents the total extra cost
 $\overline {\cal K}$ that should  be
considered to adjust the option price when the risk aversion coefficient
$\lambda = 1$.
 Notice that  $\overline {\cal K}$ has a minimun
for $\tau \simeq 10$ days (indicated by the cross) which is, for this
choice of parameters, the optimal trading time. Total extra cost at $\tau =
10$ is .77
which represents an appreciable fraction of the `risk-neutral' price.

b) Inset: Comparison between the residual risk corresponding to our optimal
strategy (full line) and the Black and Scholes strategy (dotted
line). As expected the latter is larger.

 \vfill
\eject

{\bf Inset: Technical sheet.}
\vskip 2cm

In order not to overload the main text with mathematical formulae, we give in
this separate inset the most important final equations used for our numerical
implementation, where we have neglected interest rate effects (i.e. $r \equiv
0$). The optimal strategy $\phi^*(x,t)$ is given by:

$$
\phi^*(x,t) =  \int_{x_c}^\infty dx' \langle {dx \over dt}\rangle
_{(x,t)\longrightarrow
(x',T)}
{(x'-x_c) \over {\cal D}(x)} P(x',T|x,t) \eqno(I.1)
$$
where $P(x',t'|x,t)$ is the probability that
the value of $x'$ occurs (within $dx'$) at $t'$, knowing that it was $x$ at $t
\leq t'$, $\langle {dx \over dt}\rangle _{(x,t)\longrightarrow (x',T)}$ is
the mean
instantaneous
increment conditioned to the initial condition $(x,t)$ and
 a final condition $(x',T)$, and $D(x)$ is the local volatility.
The residual risk ${\cal R}^*$ then reads:
$$
{\cal R}^* = {\cal R}_c - \sum_{t=0}^T
\int_{-\infty}^{+\infty} dx {\cal D}(x) P(x,t|x_0,0) \phi^{*2}(x,t), \eqno(I.2)
$$
where ${\cal R}_c$ is the ``bare'' risk
which would prevail in the absence of trading ($\phi(x,t) \equiv 0$):
$${\cal R}_c = \left[\int_{x_c}^{\infty} dx (x-x_c)^2
P(x,T|x_0,0) \right]- \left[\int_{x_c}^{\infty} dx (x-x_c)
P(x,T|x_0,0) \right]^2 \eqno(I.3)$$.
Finally, the option price is determined using Eq. (2) as:

$${\cal C}(x_0,x_c,T) =  \left[ \int_{x_c}^{+\infty} dx'
P(x',T|x_0,0) (x'-x_c) + \lambda \sqrt{\cal R^*}\right]\eqno(I.4)$$

which allows one to obtain numerically the option price, once $P(x',t'|x,t)$ is
reconstructed, which we do using the histogram of daily variations plus the
assumption that these daily increments are uncorrelated.

\vfill
\eject

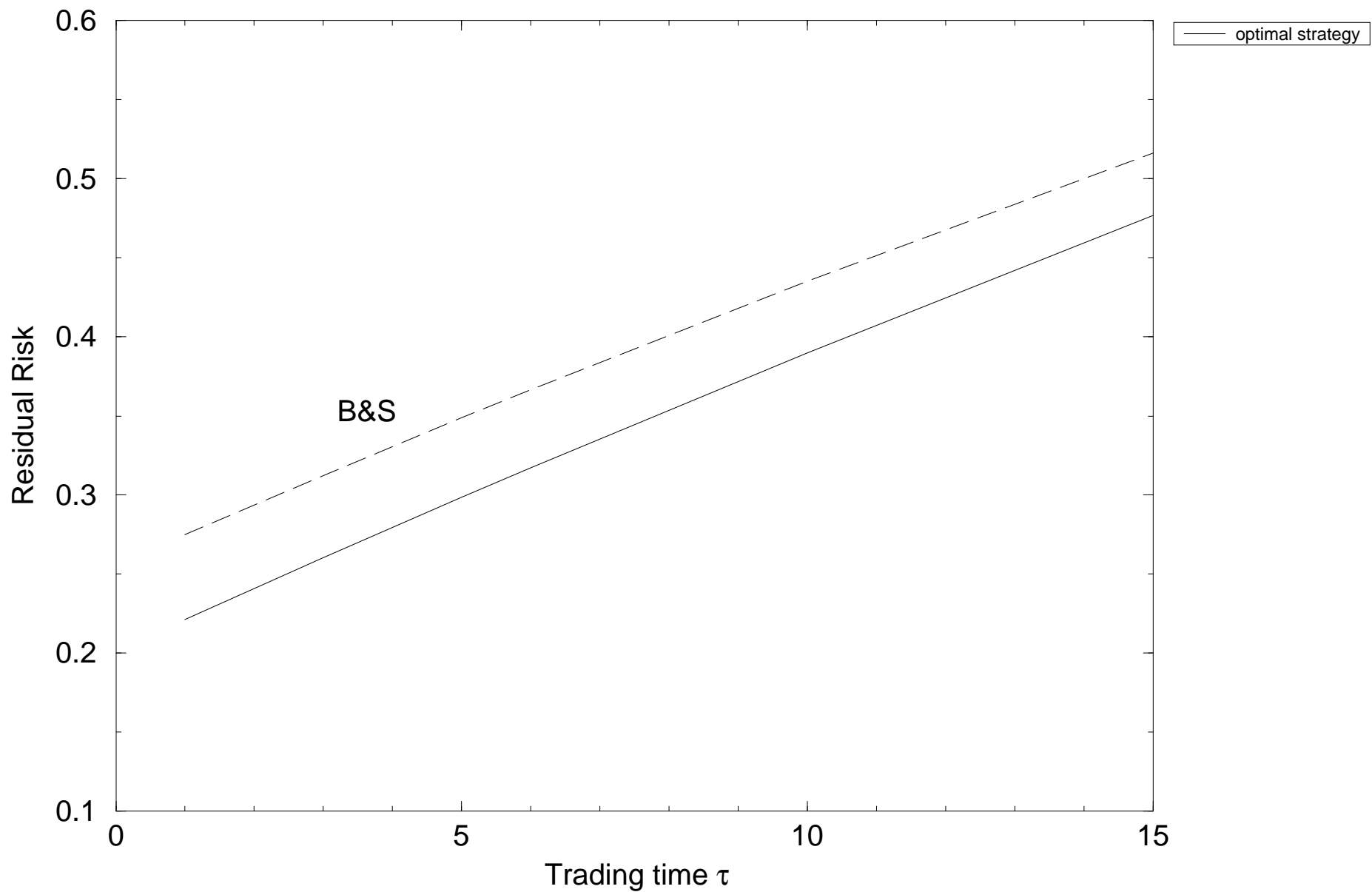
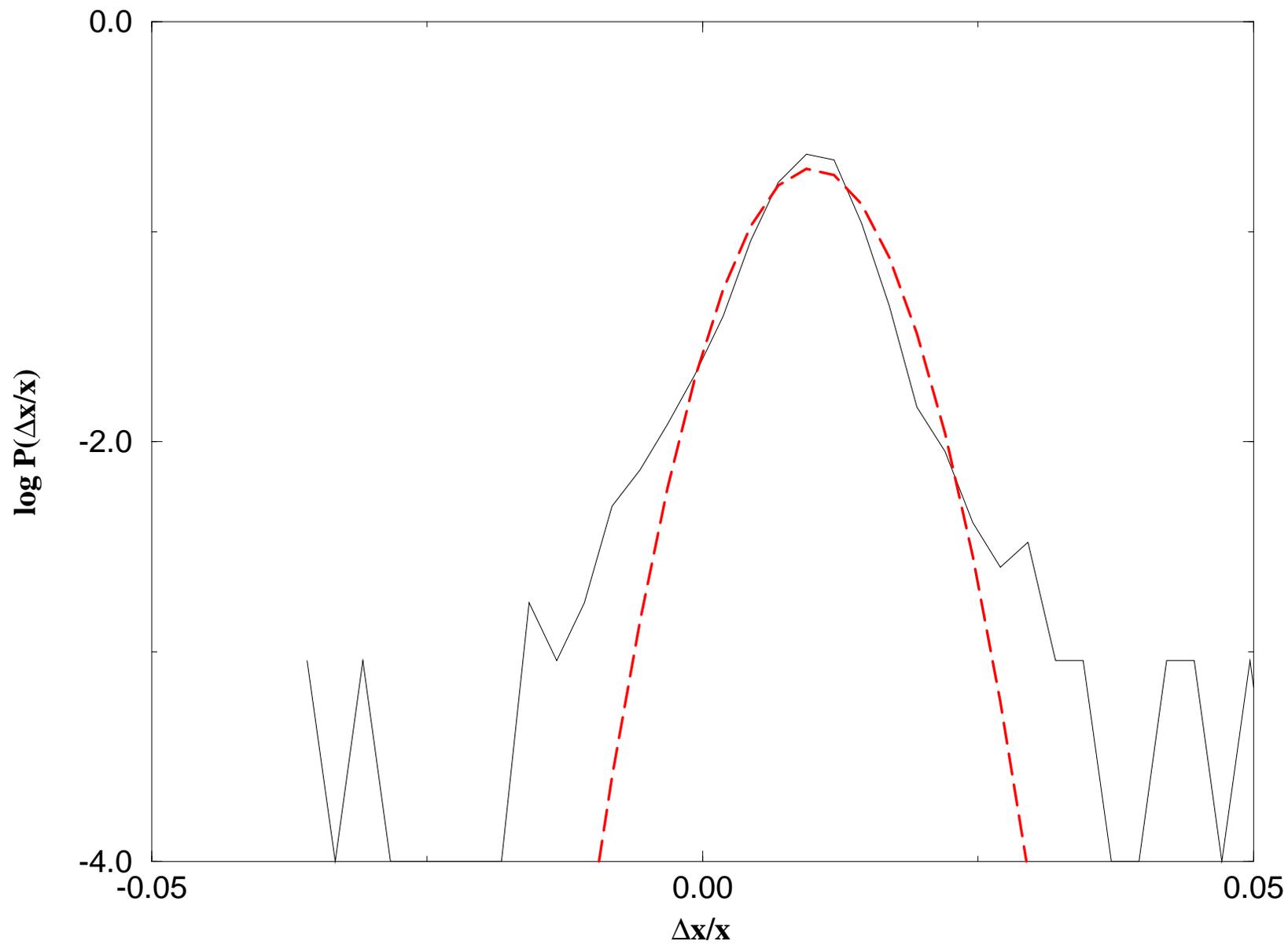
\end{document}